\documentclass[12pt,preprint]{aastex}

\slugcomment{
submitted to ApJL
}

\shorttitle{Prompt Emission Model for GRB~980425}
\shortauthors{Yamazaki et al.}

\newcommand{\beq}{\begin{equation}}
\newcommand{\beqa}{\begin{eqnarray}}
\newcommand{\eeq}{\end{equation}}
\newcommand{\eeqa}{\end{eqnarray}}

\begin{document}

\title{ An Off-Axis Jet Model For GRB~980425 and 
Low Energy Gamma-Ray Bursts}

%\author{Ryo Yamazaki\altaffilmark{1}}
%\affil{Department of Physics, Kyoto University, Kyoto 606-8502, Japan}
%\email{yamazaki@tap.scphys.kyoto-u.ac.jp}
%\and
%\author{Daisuke Yonetoku\altaffilmark{2}}
%\affil{Department of Physics, Kanazawa University, Kakuma,
%Kanazawa, Ishikawa 920-1192, Japan}
%\email{yonetoku@astro.s.kanazawa-u.ac.jp}
%\and
%\author{Takashi Nakamura\altaffilmark{3}}
%\affil{Department of Physics, Kyoto University, Kyoto 606-8502, Japan}
%\email{takashi@tap.scphys.kyoto-u.ac.jp}
%\and
%\author{Kunihito Ioka\altaffilmark{3}}
%\affil{Department of Earth and Space Science, Osaka University, 
%Toyonaka 560-0043, Japan}
%\email{ioka@vega.ess.sci.osaka-u.ac.jp}

\author{Ryo~Yamazaki\altaffilmark{1}, 
Daisuke~Yonetoku\altaffilmark{2},
and Takashi~Nakamura\altaffilmark{1}}
\altaffiltext{1}{Department of Physics, 
Kyoto University, Kyoto 606-8502, Japan}
\altaffiltext{2}{Department of Physics, Kanazawa University, 
Kakuma, Kanazawa, Ishikawa 920-1192, Japan}
\email{
yamazaki@tap.scphys.kyoto-u.ac.jp,
yonetoku@astro.s.kanazawa-u.ac.jp,
takashi@tap.scphys.kyoto-u.ac.jp
}

\def\E{{\cal E}}
\def\d{{\rm d}}
\def\p{\partial}
\def\w{\wedge}
\def\o{\otimes}
\def\f{\frac}
\def\tr{{\rm tr}}
\def\Half{\frac{1}{2}}
\def\half{{\scriptstyle \frac{1}{2}}}
\def\T{\tilde}
\def\RA{\rightarrow}
\def\N{\nonumber}
\def\n{\nabla}
\def\bb{\bibitem}
\def\BE{\begin{equation}}
\def\EE{\end{equation}}
\def\BEA{\begin{eqnarray}}
\def\EEA{\end{eqnarray}}
\def\L{\label}
\def\VVM{\langle V/V_{\rm max}\rangle}
\def\zero{{\scriptscriptstyle 0}}
\begin{abstract}
Using a simple off-axis jet model of GRBs,
we can  reproduce the observed unusual properties of the prompt
emission of GRB~980425, such as the extremely low isotropic 
equivalent $\gamma$-ray energy, the low peak energy, 
the high fluence ratio, and the long  spectral lag when 
the jet with the standard energy of $\sim10^{51}$~ergs
and the opening half-angle of 
$10\degr\lesssim\Delta\theta\lesssim 30\degr$
is seen from the off-axis viewing angle
$\theta_v\sim \Delta \theta +10\gamma^{-1}$,
where $\gamma$ is a Lorentz factor of the jet.
For our adopted fiducial parameters,
if the jet that caused GRB~980425 is viewed from 
the on-axis direction,
the intrinsic peak energy $E_p(1+z)$ is $\sim$2.0--4.0~MeV,
which corresponds to those of GRB~990123 and GRB~021004.
We also discuss the connection of GRB~980425 in our model 
with the X-ray flash, and
the origin of a class of GRBs with small $E_\gamma$.
\end{abstract}


%
%\newpage
\section{INTRODUCTION}

Recently, a very luminous gamma-ray burst (GRB),
 GRB~030329 at the distance of 0.8~Gpc ($z=0.1685$)
was confirmed to  be associated with supernova SN~2003dh 
(Stanek et al. 2003; Uemura et al. 2003; Price et al. 2003;
Hjorth et al. 2003).
The geometrically corrected $\gamma$-ray energy $E_\gamma$
of this event $\sim 5 \times 10^{49}$ergs is a factor
20  smaller than the standard value, 
if the jet break time of $\sim0.48$ days 
is assumed (Vanderspek et al. 2003; Price et al. 2003). 
GRB~980425 was the  first GRB associated with 
a supernova (SN) event, SN~1998bw at $z=0.0085$ (36~Mpc)
(Galama et al 1998; Kulkarni et al. 1998; 
Woosley et al. 1998; Pian et al 2000, 2003).
There are some other events that might 
be associated with supernovae (Della~Valle et al. 2003;
Wang \& Wheeler 1998; Germany et al. 2000; Rigon et al. 2003).
Therefore the association of  the long duration  GRBs with 
supernovae is strongly suggested and at least some
GRBs arise from the  collapse of a massive star.

In this context, it is important to investigate whether 
 GRB~980425/SN~1998bw is similar to more or less 
typical long duration  GRBs like GRB~030329/SN~2003dh.
However, GRB~980425 showed unusual observational properties.
The isotropic equivalent $\gamma$-ray energy
is $E_{iso}\sim 6\times10^{47}$~ergs and
the geometrically corrected energy is 
$E_\gamma\sim3\times10^{46}$~ergs~$(\Delta\theta/0.3)^2$, where
$\Delta\theta$ is the unknown  jet opening half-angle.
These energies are much smaller than the typical values of  GRBs 
 $E_\gamma\sim 1\times10^{51}$~ergs
(Bloom et al. 2003; Frail et al. 2001).
Bloom et al. (2003) claim that there should be some events
with small $E_\gamma$ such as GRB~980519 and GRB~980326
and  that GRB~980425 might be a member of this class.
The other properties of GRB~980425 are also unusual;
the large low energy flux, the long spectral lag, 
the low variability,
and the slowly decaying X-ray luminosity of its counterpart
detected and monitored by BeppoSAX and by XMM-Newton
(Frontera et al. 2000a; Norris, Marani, \& Bonnell, 2000; 
Fenimore \& Ramirez-Ruiz 2000; Pian et al. 2000, 2003).

Previous works suggest that the above peculiar observed properties 
of GRB~980425 might be explained
if the standard jet is seen from the off-axis viewing angle
(Ioka \& Nakamura 2001; Nakamura 1999; Nakamura 2001;
see also Maeda et al. 2002; Iwamoto 1999;
Dado, Dar, \& De~R\'{u}jula 2003; Dar \& De~R\'{u}jula 2000).
Following this scenario, the relativistic beaming effect 
reduces $E_{iso}$ and hence $E_\gamma$.
The quantity $E_{iso}$ is roughly proportional to 
$\delta^{2\sim 3}$ for
the typical observed  spectrum,
where $\delta=[\gamma(1-\beta\cos(\theta_v-\Delta\theta))]^{-1}$
is the Doppler factor and $\theta_v$ is the viewing angle
(Yamazaki, Ioka, \& Nakamura 2002).
Since $E_{iso}$ is $\sim10^{4\sim 5}$ times smaller than the 
standard value, $\delta$ should be $20 \sim 10^2$
 times smaller than the usual value.
%
%However, 
Then the peak energy $E_p$ ($\propto\delta$) becomes 
$20\sim10^2$ times smaller than  on-axis $E_p$,
that is measured when the jet is seen from the on-axis viewing angle.
However,
the observed $E_p$ of GRB~980425 ($\sim 50$~keV) is only 
a factor 4 or 5 smaller than the typical value of $\sim250$~keV. 
Therefore, one might consider that GRB~980425 belongs to a different
class of GRBs.

It is well known that the distribution of $E_p$ is log-normal with  
the mean of $\langle E_p\rangle\sim250$~keV (Preece et al. 2000).
Ioka \& Nakamura (2002) showed that if the distribution of
intrinsic $E_p$ (i.e. $E_p(1+z)$) is  log-normal, 
the redshifted one is also log-normal under the assumption that 
the redshifts of the observed GRBs are random. 
Therefore,  $\langle E_p(1+z)\rangle\sim570$~keV since the mean value
of the measured redshifts is $\sim 1.3$ (Bloom et al. 2003). 
There are some GRBs with even higher intrinsic peak energy; 
for example, $E_p(1+z)\sim2.0$~MeV for GRB~990123 (Amati et al. 2002) 
and $E_p(1+z)\sim3.6$~MeV for GRB~021004 (Barraud  et al. 2003).
Furthermore, Fig.~3 of Schaefer et al. (2003) shows that the 
highest value of $E_p(1+z)$ detected by BATSE is about 4~MeV. 
Since  GRB~980425 is the nearest GRB,
the redshift factor is not important. 
In this sense, the peak energy of GRB~980425 is at least a factor
$\sim$ 10 smaller than the usual one of $\sim570$~keV.
Suppose that the intrinsic $E_p$ of GRB~980425 is similar to
that of GRB~990123 and GRB~021004
when the jet of GRB~980425 is seen from the 
on-axis viewing angle.
Then, the observed $E_p$ of GRB~980425 is 
$\sim10^2$ times smaller than the intrinsic $E_p$ of
 GRB~990123 and GRB~021004.
%
%If the on-axis peak energy of GRB980425 is similar to
% GRB990123 and GRB021004, the observed peak energy is
%a factor 40 to 80 smaller than the intrinsic $E_p$ of
% GRB990123 and GRB021004 so that
%this smallness can be 
%compared with that of  $E_\gamma$.
This is the reason why  we incline to reconsider 
 the off-axis jet model for GRB~980425. 

In this Letter, assuming the  rather large on-axis $E_p$, 
we reconsider the prompt emission of GRB~980425
using the simple model in  Yamazaki, Ioka, \& Nakamura (2002, 2003b)
to reproduce its unusual observed quantities.
In \S~\ref{sec:analysis}, 
in order to extract the observational properties that should be 
compared with our theoretical model for prompt emission of the GRB,
we analyze the BATSE data of GRB~980425. 
In \S~\ref{sec:model}, we describe a simple jet model 
including the cosmological effect.
We assume a uniform jet with a sharp edge. 
Numerical results are shown in \S~\ref{sec:results}.
Section~\ref{sec:dis} is devoted to  discussions.
Throughout this paper, we adopt the flat universe with 
$\Omega_M=0.3$, $\Omega_\Lambda=0.7$ and $h=0.7$.

%
%\newpage

\section{SPECTRAL ANALYSIS FOR PROMPT EMISSION OF
GRB~980425 USING BATSE DATA}\label{sec:analysis}
In our  simple jet model of GRBs,
% where instantaneous emission of 
%infinitesimally thin shell is assumed,  
the time-dependence of spectral indices is not treated,
while it is known that the spectral parameters of GRB~980425 changed
in time (Galama et al. 1998; Frontera et al. 2000a).
Hence, we should discuss the time-averaged observed spectral 
properties of GRB~980425 before we apply our model to them.

Using the BATSE data of GRB~980425,
we analyze the spectrum within the time of Full-Width at 
Half-Maximum of the peak flux in the light curve of 
BATSE channel~2 (50--110 keV).
This time interval approximately corresponds to portions
 ``B'' and ``C'' in Frontera et al. (2000a),
when most of photons arrived at the detector
and  the spectral shape was approximately constant with time.
%
%We fit the observed spectrum with smoothly broken power-law function
%(Band et al. 1993) given by
%\begin{eqnarray}
%N(E) = \left\{ 
%\begin{array}{ll}
%A \Bigl( \frac{E}{100~{\rm keV}} \Bigr)^{\alpha}
%\exp(- \frac{E}{E_{0}}) 
%& {\rm for} \ E \le (\alpha - \beta) E_{0}\\
%A \Bigl( \frac{E}{100~{\rm keV}} \Bigr)^{\beta} \Bigl( 
%\frac{(\alpha - \beta) E_{0}}{100~{\rm keV}}\Bigr)
%^{\alpha - \beta} \exp(\beta - \alpha) 
%& {\rm for} \ E \ge (\alpha - \beta) E_{0}
%\end{array}\right.
%\end{eqnarray}
%where $N(E)$ is in unit of ${\rm photons~cm^{-2}~s^{-1}~keV^{-1}}$,
%$E_{0}$ is the energy at the spectral break, and $\alpha$ and $\beta$ 
%are the low- and high-energy power low index, respectively.
%
We fit the observed spectrum with smoothly broken power-law function
given by Band et al. (1993), that is characterized by the 
energy at the spectral break $E_0$, and
the low- and high-energy photon index $\alpha$ and $\beta$,
respectively.
For the case of $\beta < -2$, the peak energy is derived 
as $E_{p} = (2+\alpha) E_{0}$. The best-fit spectral parameters are
\begin{eqnarray}
&& \alpha = -1.0 \pm 0.3 \N\\
&& \beta  = -2.1 \pm 0.1 \N\\
&& E_{p} = 54.6 \pm 20.9 \ {\rm keV} \N.
\end{eqnarray}
The reduced chi square is 1.10 for 31 degree of freedom. 
These results are consistent with those derived by the previous works 
(Frontera et al. 2000a; Galama et al. 1998).
Although the photon indices are the typical values of GRBs,
$E_p$ is lower than the typical values of GRBs (Preece et al. 2000). 
This spectral property is similar to one of the
recently identified class of the X-ray flash 
(Kippen et al. 2002; Heise et al. 2001). 

The observed fluence of the entire emission between 20 to 2000 keV is
$S(20-2000~{\rm keV}) = (4.0\pm0.74)\times10^{-6}~{\rm erg~cm^{-2}}$,
so the isotropic equivalent $\gamma$-ray energy becomes 
$E_{iso} = (6.4\pm1.2)\times10^{47}$~ergs. 
The fluence ratio is 
$R_s = S(20-50\ {\rm keV})/S(50-320\ {\rm keV})=0.34 \pm 0.036$.
In the following sections, we reproduce the above results
using our prompt emission model.

\section{MODEL OF PROMPT EMISSION OF GRBs}\label{sec:model}
We use a simple jet model of prompt emission of GRBs
adopted in Yamazaki et al. (2003b), where 
the cosmological effect is included
(see also Yamazaki et al. 2002, 2003a; Ioka \& Nakamura 2001).
We adopt an instantaneous emission of infinitesimally thin shell 
at $t=t_0$ and $r=r_0$.
Then the observed flux of a single pulse is given by
\begin{eqnarray}
F_{\nu}(T)
=\f{2(1+z)r_0 c A_0}{d_L^2}
{{\Delta \phi(T) f\left[\nu_z\gamma (1-\beta\cos\theta(T))\right]
}\over{\left[\gamma (1-\beta\cos\theta(T))\right]^2}},
\label{eq:jetthin}
\end{eqnarray}
where, $1-\beta\cos\theta(T)=(1+z)^{-1}({c\beta}/{r_0})(T-T_0)$
%and $T_0=(1+z)(t_0-r_0/c\beta)$,
and  $A_0$ determines the normalization of the emissivity.
Detailed derivation of Eq.~(\ref{eq:jetthin}) and the 
definition of $\Delta\phi(T)$
are found in Yamazaki et al. (2003b).
In order to have a spectral shape similar to that derived by 
the previous section,
we adopt the following form of the spectrum in the comoving frame,
\begin{equation}
f(\nu')= \left\{ 
\begin{array}{ll}
  (\nu'/\nu'_0)^{1+\alpha_B}\exp(-\nu'/\nu'_0)
         & {\rm for} \ \nu'/\nu'_0 \le \alpha_B - \beta_B \\
  (\nu'/\nu'_0)^{1+\beta_B}(\alpha_B-\beta_B)^{\alpha_B-\beta_B}
        \exp(\beta_B-\alpha_B)
         & {\rm for} \ \nu'/\nu'_0 \ge \alpha_B - \beta_B \ ,
\end{array}
\right.
\label{eq:spectrum}
\end{equation}
with $\alpha_B=-1$ and $\beta_B=-2.1$.
Equations (\ref{eq:jetthin}) and (\ref{eq:spectrum})
are the basic equations to calculate the flux of a single pulse,
which depends on the following parameters;
$\gamma$, $\gamma \nu_0'$, $\theta_v$, $\Delta \theta$,
$r_0/c \beta \gamma^2$, $z$, and $A_0$.
In the next section, the viewing angle $\theta_v$ and the jet
opening half angle $\Delta\theta$ are mainly varied.
The other parameters are fixed as follows;
the quantity $\gamma$ is fixed as $\gamma=100$.
The isotropic $\gamma$-ray energy is calculated as
$E_{iso}=4\pi(1+z)^{-1}d_L^2 S(20-2000\,{\rm keV})$,
where 
$S(\nu_1-\nu_2)$
is the observed fluence in the energy range $h\nu_1$--$h\nu_2$~keV.
We fix the amplitude $A_0$ so that the geometrically-corrected 
$\gamma$-ray energy $E_\gamma=(\Delta\theta)^2E_{iso}/2$ be
observationally preferred value
when we see the jet from the on-axis viewing angle $\theta_v=0$.
It is shown that $E_\gamma$ is tightly clustering about
a standard energy $\E_\gamma$ of
 $\sim10^{51}$~ergs (Bloom et al. 2003; see also Frail et al. 2001;
Panaitescu \& Kumar 2002).
Bloom et al. (2003) derived this energy as
\begin{equation}
\log \E_\gamma=\log\left[1.15\times10^{51}(h/0.7)^{-2}\
{\rm ergs}\right] \pm0.07 \ ,
\end{equation}
so that $\E_\gamma=$(0.98--1.35)$\times10^{51}$~ergs, at the 
1~$\sigma$ level while  $\E_\gamma=$(0.51--2.57)$\times10^{51}$~ergs, 
at 5 $\sigma$ level.
Note that the smaller jet opening half-angle $\Delta\theta$
corresponds to the larger $A_0$ (Yamazaki et al. 2003b).

Practical calculations show that
when the jet with $\alpha_B=-1$ and $\beta_B=-2.1$ is
seen from the on-axis viewing angle $\theta_v=0$,
the observed peak energy becomes 
$E_p^{(\theta_v=0)}\sim1.54\,\gamma\nu'_0(1+z)^{-1}$,
which is independent on 
$\Delta\theta$ larger than $\sim\gamma^{-1}$.
In order to reproduce the observed quantities of GRB~980425,
we adopt the value $\gamma\nu'_0=2600\,{\rm keV}$, 
which yields $E_p^{(\theta_v=0)}(1+z)\sim 4.0$~MeV.
For comparison, we consider another case of
$\gamma\nu'_0=1300$~keV, which reads
$E_p^{(\theta_v=0)}(1+z)\sim2.0$~MeV.
These values  correspond to the intrinsic $E_p$ of
GRB~021004 and GRB~990123, respectively.
Note here that in our jet model 
the quantities that will be calculated
in the next section do not depend on $r_0/c\beta \gamma^2$;
for example, 
$E_{iso}\propto A_0(r_0/c\beta \gamma^2)^2\propto
(r_0/c\beta \gamma^2)^0$ since
$A_0\propto(r_0/c\beta \gamma^2)^{-2}$.
The value of $r_0/c\beta \gamma^2$ will be determined
when we discuss the spectral lag in \S~\ref{sec:dis}.

\section{ISOTROPIC ENERGY, PEAK ENERGY, AND FLUENCE RATIO}
\label{sec:results}

We now calculate the isotropic equivalent 
$\gamma$-ray energy $E_{iso}$ as a function of $\theta_v$
and $\Delta\theta$. 
Then, the peak energy $E_p$ and the fluence ratio 
$R_s=S(20-50 \ {\rm keV})/S(50-320 \ {\rm keV})$ 
are computed for the set of
$\Delta\theta$ and $\theta_v$ that reproduces the observed
$E_{iso}$ of GRB~980425.

For fixed $\Delta\theta$ and $\E_\gamma$, 
$E_{iso}$ is calculated as a function of the viewing angle $\theta_v$.
The result is shown in Fig.~1.
When $\theta_v\lesssim\Delta\theta$,
  $E_{iso}$ is essentially constant, while
for $\theta_v\gtrsim\Delta\theta$,
$E_{iso}$ is considerably smaller than the typical value of
$\sim10^{51-53}$~ergs because of the relativistic beaming effect.
In order to explain the observation,
$\theta_v$ should be $\sim21\degr$ in the case of 
$\Delta\theta=15\degr$, while 
$\theta_v\sim25\degr$  in the case of $\Delta\theta=20\degr$.
This result does not  depend on $\gamma\nu'_0$ so much.

The upper panels of Fig.~2 and 3 show $\theta_v^\ast$, 
for which $E_{iso}$ becomes equal to the observed values, 
as a function of $\Delta\theta$
in the case of $\gamma\nu'_0=2600$~keV and 1300~keV, respectively.
Since the emissivity ($\propto A_0$) of the jet
is small for large $\Delta\theta$,
the relativistic beaming effect should be weak for large
$\Delta\theta$.
Therefore, the value of $\theta_v^\ast-\Delta\theta$ is a 
decreasing function of $\Delta\theta$.
For such $\theta_v^\ast$, we calculate 
the fluence ratio $R_s^\ast=R_s^{(\theta_v=\theta_v^\ast)}$
and the peak energy $E_p^\ast=E_p^{(\theta_v=\theta_v^\ast)}$.
The middle and the lower panels of Figs.~2 and 3 show the results.
The quantity $E_p^\ast$ is proportional to the Doppler factor
%$\delta\sim2\gamma[1+(\gamma(\theta_v^\ast-\Delta\theta))^2]^{-1}$.
$\delta\sim[\gamma(1-\beta\cos(\theta_v^\ast-\Delta\theta))]^{-1}$.
Therefore, when $\Delta\theta$ increases, 
$\theta_v^\ast-\Delta\theta$ decreases so that $E_p^\ast$ increases.
Since we fix spectral indices $\alpha_B$ and $\beta_B$,
$R_s^\ast$ depends only on $E_p^\ast$.
Hence, if $E_p^\ast$ is large,  the spectrum is hard and 
 $R_s^\ast$ is small.
%
%Since we see the jet with the off-axis viewing angle 
%$\theta_v>\Delta\theta$, spectral hardness becomes soft since 
%observed $E_p$ shifts toward low energy side due to the
%relativistic Doppler effect (Yamazaki et al. 2002).
%
For the fiducial parameters of 
$\gamma\nu'_0=2600$~keV,
$\E_\gamma=1.15\times10^{51}$~ergs,
and $E_{iso}=6.4\times10^{47}$~ergs,
$\Delta\theta$ should be  between $\sim18\degr$ and 
$\sim31\degr$,
and then $\theta_v^\ast$ ranges between $\sim24\degr$ 
and $\sim35\degr$
in order to reproduce the observed values of $R_s$ and $E_p$.
When $\E_\gamma$ is varied from $0.51\times10^{51}$ 
to $2.6\times10^{51}$~ergs (at 5~$\sigma$ level), the 
allowed region with $20\degr\lesssim\Delta\theta\lesssim30\degr$
can exist even in the case of $\gamma\nu'_0=1300$~keV.
 
Note that $\gamma$ does not affect our results for
observed $R_s^\ast$ and $E_p^\ast$.
When $\gamma$ is large, $\theta_v^\ast$ becomes small
because the observed flux for fixed $\theta_v$ becomes small due to
stronger relativistic beaming effect.
However, we can see that $\gamma(\theta_v^\ast-\Delta\theta)$
 remains almost unchanged even if $\gamma$ is varied.
Then for fixed $\gamma\nu'_0$, 
$E_p^\ast$ remains constant  since 
$E_p^\ast\propto\nu'_0\delta
\sim2\gamma\nu'_0[1+(\gamma(\theta_v^\ast-\Delta\theta))^2]^{-1}$.
%(Yamazaki et al. 2002).
The quantity $R_s^\ast$ depends only on $E_p^\ast$ 
so that $\gamma$ does not affect the estimate of $R_s^\ast$.

\section{DISCUSSION}\label{sec:dis}

We considered the time-averaged emissions, which means that
successive emissions from multiple subjets (or shells) are approximated by
one spontaneous emission caused by a single jet
(Yamazaki et al. 2002).
We choose $\alpha_B=-1$, $\beta_B=-2.1$, $\gamma=100$, and
$\gamma\nu'_0=2600$~keV for the canonical set of parameters.
As a result, when the jet of opening 
half-angle of $\Delta\theta\sim10$--30$\degr$ is seen from the 
off axis viewing angle of $\theta_v\sim\Delta\theta+6\degr$,
observed quantities can be well explained.
Derived $\theta_v$ and $\Delta \theta$ are 
consistent with those suggested  in
Nakamura 2001, Nakamura 1999, and Maeda et al. 2002.
We may also be able to  explain observed low variability
since only  subjets at the edge of the cone contribute to 
the observed quantities (see the discussion of Yamazaki et al. 2002).
If the jet is seen from an on-axis viewing angle 
(i.e. $\theta_v<\Delta\theta$),
the intrinsic peak energy $E_p(1+z)$ is $\sim4.0$~MeV,
which is almost the same as the highest one 
(Schaefer 2003; Amati et al. 2002; Barraud et al. 2003).

As we have mentioned in \S~\ref{sec:model},
$E_{iso}$, $E_p^\ast$, and $R_s^\ast$ do not
depend on the parameter $r_0/\beta c\gamma^2$.
In order to estimate the value of $r_0/\beta c\gamma^2$,
we discuss the spectral lag of GRB~980425 (Ioka \& Nakamura 2001).
In our model, we can calculate the spectral lag $\Delta T$,
which is defined, for simplicity, as the difference of the peak time
between BATSE energy channel 1 and 3.
We obtain
$\Delta T/(r_0/c\beta \gamma^2)=$0.97--1.34.
Therefore, the observed value of $\Delta T=3$~s (Norris et al. 2000)
can be explained when  $r_0/c\beta \gamma^2=(2.2$--3.1)~sec,
which is in the reasonable parameter range.

The observed quantities of small $E_p$ and large fluence ratio
$R_s$ (see also Frontera et al. 2000a) are the typical
values of the X-ray flash 
(Heise et al. 2001; Kippen et al. 2002; see also
Barraud et al. 2003; Arefiev, Priedhorsky \& Borozdin 2003).
The operational definition of the X-ray flash
detected by {\it BeppoSAX} is a fast X-ray transient with duration 
less than $\sim10^3$~seconds which is detected by WFCs 
and not detected by the GRBM (Heise et al. 2001).
If the distance to the source of GRB~980425 that has
an opening half-angle of $\Delta\theta=20\degr$ were
larger than $\sim86$~Mpc, the observed flux in the $\gamma$-ray
band would have been less than the limiting sensitivity of GRBM
$\sim5\times10^{-7}$~erg~cm$^{-2}$ in 40--700~keV band (Band 2003),
so that the event would have been detected as an X-ray flash. 

We might be able to explain the origin of a class with low 
$E_\gamma$, pointed by Bloom et al. (2003).
Let us consider the jet seen from a viewing angle
$\theta_v\sim\Delta\theta+\gamma_i^{-1}$,
where $\gamma_i$ is the Lorentz factor of a prompt $\gamma$-ray 
emitting shell.
Due to the relativistic beaming effect, observed $E_\gamma$ of such
a jet becomes an order of magnitude smaller than the standard energy
%of $\sim10^{51}$~ergs 
(see Fig.~1).
At the same time, the observed peak energy $E_p$ is small
because of the relativistic Doppler effect.
In fact, the observed $E_p$ of GRB~980326 and GRB~981226 are
$\sim35$~keV and $\sim60$~keV, respectively
(Amati et al. 2002; Frontera et al. 2000b).
In our model  the fraction of GRBs with  low  $E_\gamma$
becomes $2/(\gamma_i\Delta\theta)\sim 0.1$ since
the mean value of $\Delta \theta \sim 0.2$, 
while a few GRBs with  low  $E_\gamma$ are observed
in $\sim$ 30 samples (Bloom et al. 2003).
In later phase, the Lorentz factor of afterglow emitting shock
$\gamma_f$ is smaller than $\gamma_i$, so that
$\theta_v<\Delta\theta+\gamma_f^{-1}$.
Then, the observed properties of afterglow may be similar to  the 
on-axis case $\theta_v\ll\Delta\theta$;
hence the observational estimation of the jet break time and the 
jet opening angle remains the same.

%%% acknowledgments section 

\acknowledgments
We would like to thank the referee for useful comments and
suggestions.
Numerical computation in this work was carried out at the 
Yukawa Institute Computer Facility.
This work was supported in part by
Grant-in-Aid for Scientific Research 
of the Japanese Ministry of Education, Culture, Sports, Science
and Technology, No.05008 (R.Y.), 
No.14047212 (T.N.), and No.14204024 (T.N.).

%% Appendix material 
\appendix

%\section{}\label{}

%%% The reference list 

\clearpage

\begin{figure}
\plotone{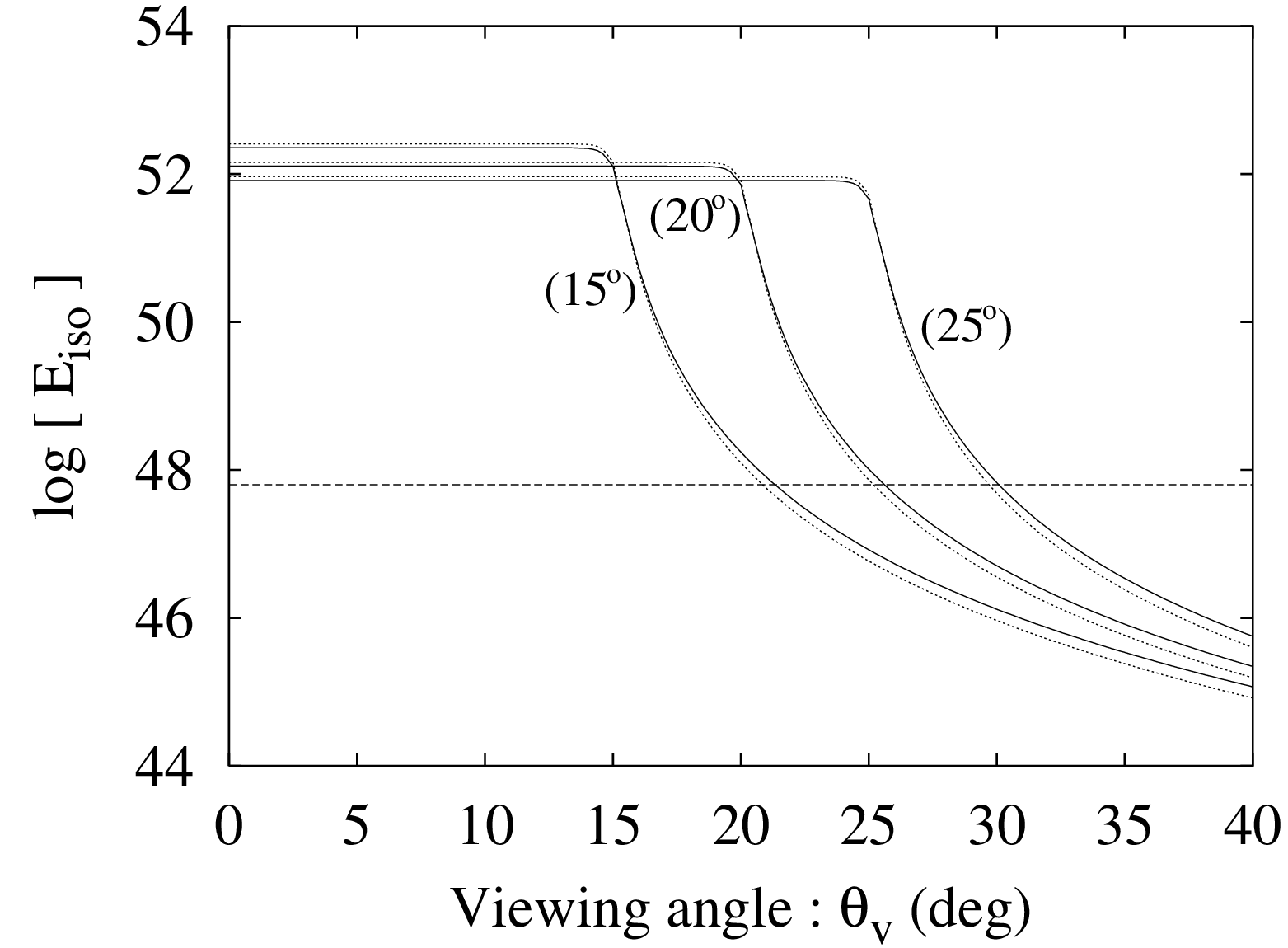}
\caption{The isotropic equivalent $\gamma$-ray energy $E_{iso}$
is shown as a function of the viewing angle $\theta_v$ for a
fixed jet opening half-angle $\Delta\theta$.
The source is located at $z=0.0085$.
The values of $\Delta\theta$ are shown in parentheses.
Solid lines correspond to the case of $\gamma\nu'_0=2600$~keV,
while dotted lines $\gamma\nu'_0=1300$~keV.
Other parameters are fixed as $\alpha_B=-1$, $\beta_B=-2.1$,
$\gamma=100$, and $\E_\gamma=1.15\times10^{51}$~ergs.
Horizontal dashed line represents the observed value of GRB~980425
$E_{iso}=6.4\times10^{47}$~ergs.
The value of $E_{iso}$ in the on-axis case $\theta_v<\Delta\theta$
is slightly smaller for $\gamma\nu'_0=2600$~keV than for 
$\gamma\nu'_0=1300$~keV. 
This is because the amplitude $A_0$ is fixed so that we should 
observe constant $E_\gamma$ from the source at $z=1$,
and the K-correction is larger for
$\gamma\nu'_0=1300$~keV than for $\gamma\nu'_0=2600$~keV.
}
\label{fig_1}
\end{figure}

\begin{figure}
\plotone{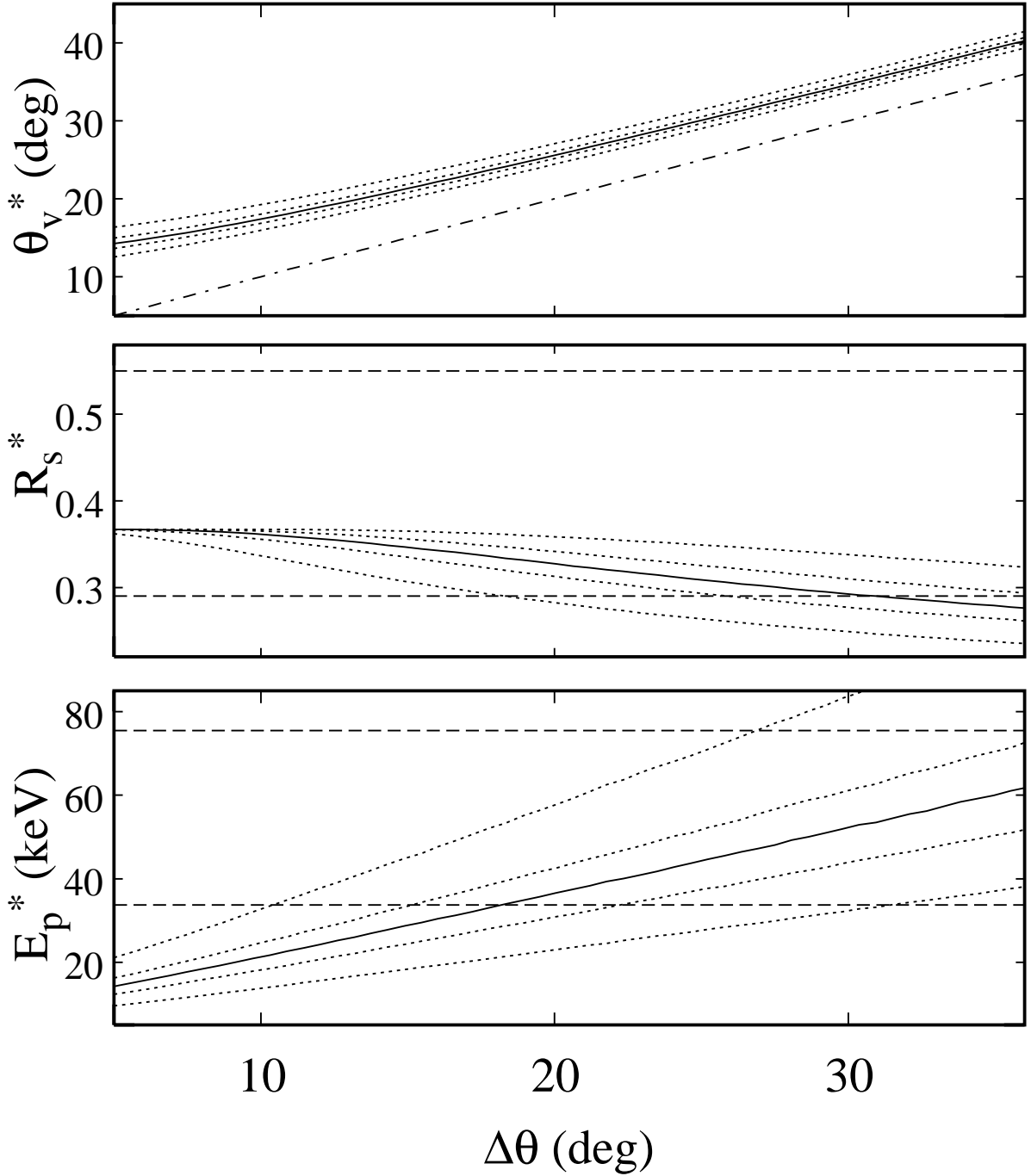}
\caption{The upper panel shows $\theta_v^\ast$ for which
$E_{iso}$ is the  observed value of GRB~980425,
while the middle and the lower panels represent the fluence ratio
$R_s^\ast=R_s^{(\theta_v=\theta_v^\ast)}$ and the peak energy
$E_p^\ast=E_p^{(\theta_v=\theta_v^\ast)}$, respectively. 
Solid lines correspond to the fiducial case of 
$E_{iso}=6.4\times10^{47}$~ergs 
and $\E_\gamma=1.15\times10^{51}$~ergs.
The dotted lines represent regions where $E_{iso}$ becomes
$(6.4\pm1.2)\times10^{47}$~ergs when
$\E_\gamma$ is in 1~$\sigma$ and 5~$\sigma$ level
around the fiducial value, respectively.
Other parameters are fixed as $\alpha_B=-1$, $\beta_B=-2.1$,
$\gamma=100$, and $\gamma\nu'_0=2600$~keV.
The dot-dashed line in the upper panel represents 
$\theta_v^\ast=\Delta\theta$.
Horizontal dashed lines in the middle and the lower panels represent 
the observational bounds $R_s=0.42\pm0.13$ and 
$E_p=54.6\pm20.9$~keV, respectively.
}
\label{fig_2}
\end{figure}

\begin{figure}
\plotone{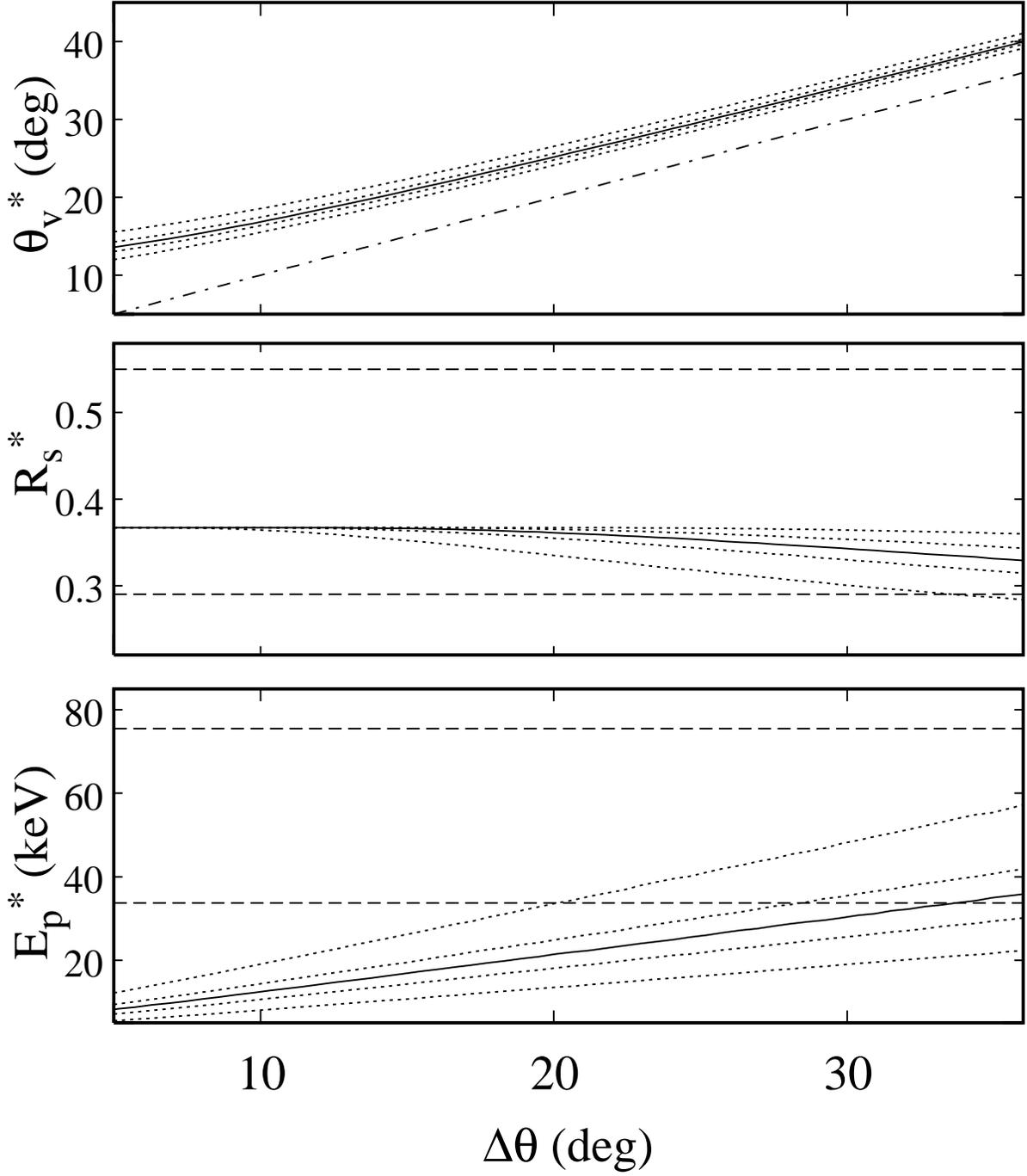}
\caption{ Same as Fig.~2 but for $\gamma\nu'_0=1300$~keV.
}
\label{fig_3}
\end{figure}

\end{document}